\documentclass[onecolumn,showpacs,preprintnumbers]{revtex4}
\usepackage{mathrsfs}
\usepackage{graphicx}
\usepackage{dcolumn}
\usepackage{bm}
\usepackage{epsfig}
\usepackage{amsmath,amssymb}
\usepackage{multirow}

\begin{document}

\title{Mass spectra of bottom-charm baryons }
\author{Zhen-Yu Li$^{1}$}
\email{zhenyvli@163.com }
\author{Guo-Liang Yu$^{2}$}
\email{yuguoliang2011@163.com }
\author{Zhi-Gang Wang$^{2}$ }
\email{zgwang@aliyun.com }
\author{Jian-Zhong Gu$^{3}$ }
\author{Hong-Tao Shen$^{4}$ }
\affiliation{$^1$ School of Physics and Electronic Science, Guizhou Education University, Guiyang 550018,
China\\$^2$ Department of Mathematics and Physics, North China Electric Power University, Baoding 071003,
China\\$^3$ China Institute of Atomic Energy, Beijing 102413,
China\\$^4$ Guangxi Key Laboratory of Nuclear Physics and Technology, Guangxi Normal University, Guilin 541006,
China}
\date{\today }

\begin{abstract}
In this paper, we investigate the mass spectra of bottom-charm baryons systematically, where the relativistic quark model and the infinitesimally shifted Gaussian basis function method are employed. Our calculation shows that the $\rho$-mode appears lower in energy than the other excited modes. According to this feature, the allowed quantum states are selected and a systematic study of the mass spectra for $\Xi_{bc}^{'}$ ($\Xi_{bc}$)  and $\Omega_{bc}^{'}$ ($\Omega_{bc}$) families is performed. The root mean square radii and quark radial probability density distributions of these baryons are analyzed as well. Next, the Regge trajectories in the $(J,M^{2})$ plane are successfully constructed based on the mass spectra. At last, we present the structures of the mass spectra, and analyze the difficulty and opportunity in searching for the ground states of bottom-charm baryons in experiment.

Key words: Bottom-charm baryons, Mass spectra, Relativistic quark  model.
\end{abstract}

\pacs{ 13.25.Ft; 14.40.Lb }

\maketitle

\section*{I. Introduction}

\label{sec1}
Quantum chromodynamics (QCD) predicts the existence of baryons with two heavy quarks ($b,c$) and one light quark, known as the doubly heavy baryons (DHBs). DHBs can be divided into three groups: double-charm baryons ($\Xi_{cc}$ and $\Omega_{cc}$), double-bottom baryons ($\Xi_{bb}$ and $\Omega_{bb}$), and bottom-charm baryons ($\Xi_{bc}$ and $\Omega_{bc}$). The study of DHBS contributes to an in-depth understanding of the heavy quark symmetry, chiral dynamics, fundamental theory of the strong interaction, and models inspired by QCD.

The experimental research of DHBs was full of twists and turns.
The $\Xi_{cc}^{+}$ baryon was first observed by the SELEX collaboration in 2002~\cite{a01}. But, other experiments have failed to confirm this so far. The experimental turning point for DHBS research occurred in 2017.
Observations of $\Xi_{cc}^{++}$ baryons were reported by the LHCb collaboration that year and have been confirmed several times since~\cite{a02,a03,a04}. Now, the $\Xi_{cc}^{++}$ baryon has become the first DHB collected in the new PDG data~\cite{a05}. It opened the door to the experimental detection of DHBs, and people expected to find more DHBs experimentally. However, in the next experiments of searching for the $\Xi_{bc}^{0}$~\cite{a06}, $\Omega_{bc}^{0}$~\cite{a07} and $\Xi_{bc}^{+}$~\cite{a08} baryons, the LHCb collaboration observed no significant signals in the invariant mass range of 6.7 $\sim$ 7.3 GeV. So far, the DHB containing bottom quark has not yet been discovered experimentally.

The zero result of searching for the bottom-charm baryons has aroused great concern. Many theoretical efforts have been carried out to predict the production in the collider~\cite{a101,a102,a103,a104,a105,a106,a107,a108}, the decay properties~\cite{a109,a1010,a1011,a1012,a1013,a1014,a1015,a1016,a1017,a1018,a1019,a1020,a1021} and the accurate mass spectrum~\cite{a1033,a1034,a1035,a1036,a1037,a1038,a1039,a1040,a1041,a1042,a1043,a1044,a1045,a1046,a1047,a1048,a1049,a1050,a1051,a1052,a1053,a1022,a1023,a1024,a1025,a1026,a1027,a10271,a10272,a1028,a1029,a1030,a1031,a1032} of DHBs, so as to provide more powerful theoretical supports to the related experiments in the near future.

Recently, we have developed an approximate method~\cite{a021,a022} to analyze the singly heavy baryon spectra systematically in which the relativistic quark model~\cite{a091,a0912} and the infinitesimally shifted Gaussian (ISG) basis function method~\cite{a092} are employed. The result indicates that this method is reasonable and effective in the singly heavy baryon spectroscopy. Additionally, it is worth pointing out that these calculations are based on a uniform set of parameters~\cite{a021,a022}. Then, we extended this method to study the mass spectra of the double charm baryons ($\Xi_{cc}$ and $\Omega_{cc}$)~\cite{a1055} and the double bottom baryons ($\Xi_{bb}$ and $\Omega_{bb}$)~\cite{a1056}. In the following sections, the systematic calculation of the bottom-charm baryon spectra by this method will be performed.

This paper is organized as follows. In
Sect.II, the calculation method used in this work is briefly introduced. In Sect.III, we present the calculation results of the bottom-charm baryons, including the root mean square (r.m.s.) radii, mass spectra, quark radial probability density distributions, Regge trajectories and spectral structure, and give discussions about the results. Sect.IV is reserved for our conclusions.

\section*{II. Phenomenological methods adopted in this work}

In our calculations, the relativistic quark model is employed to investigate the full mass spectra. In order to improve the computational accuracy and efficiency, the ISG method is also adopted in our studies. The relevant technical details can be found in references~\cite{a021,a022,a091,a0912,a092}. Next, we mainly introduce the selection of Jacobi coordinates and the structure of functions for bottom-charm baryons.
The DHB is regarded as a three-quark system and the related calculation is performed in the Jacobi coordinates as shown in Fig.1. There are three channels of Jacobi coordinates,
which are defined as
\begin{eqnarray}
&\boldsymbol\rho_{i}=\textbf{r}_{j}-\textbf{r}_{k}, \\
&\boldsymbol\lambda_{i}=\textbf{r}_{i}-\frac{m_{j}\textbf{r}_{j}+m_{k}\textbf{r}_{k}}{m_{j}+m_{k}},
\end{eqnarray}
where $i$, $j$, $k$ = 1, 2, 3 (or replace their positions in turn). $\textbf{r}_{i}$ and $m_{i}$ stand for the position vector and the mass of the $i$th quark, respectively.

\begin{figure}[htbp]
\begin{center}
\includegraphics[width=0.8\textwidth]{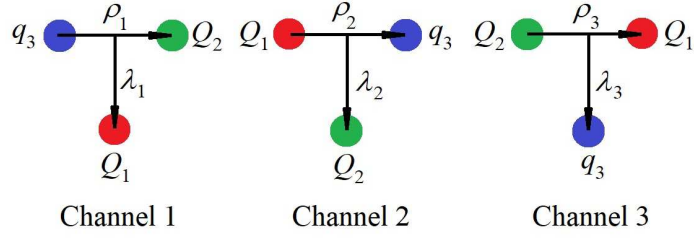}
\end{center}
\caption{(Color online)Jacobi coordinates for the three-quark system. The 3rd particle is specified as the light quark in the case of DHBs. }
\end{figure}

In our previous study of the double-bottom or -charm baryons~\cite{a1055,a1056}, the calculation result shows that the two heavy quarks in a baryon are bonded tightly. So, these two heavy quarks can be described as a subsystem, and channel 3 is a good picture to describe this situation. Similarly,
 the calculations in this work are also based on channel 3, where the 3rd quark is specified as the light quark and the flavor wave function in the subspace of the two heavy quarks ($bc$) can be expressed as,
 \begin{eqnarray}
\begin{aligned}
\Xi_{bc}=\frac{1}{\sqrt{2}}(bc-cb)q,\\
\Xi_{bc}^{'}=\frac{1}{\sqrt{2}}(bc+cb)q.
\end{aligned}
\end{eqnarray}
Here $q$ denotes up quark ($u$) or down quark ($d$). For $\Omega_{bc}$ and $\Omega_{bc}^{'}$, $q$ is replaced by strange quark.

 In channel 3, $\textbf{\emph{l}}_{\rho3}$ (denoted in short as $\textbf{\emph{l}}_{\rho}$) stands for the orbital angular momentum between the two heavy quarks, and  $\textbf{\emph{l}}_{\lambda3}$ (denoted in short as $\textbf{\emph{l}}_{\lambda}$) represents the one between the bottom-charm quark pair and the light quark.
For a quantum state in this work, the spatial wave function is combined with the spin function as follow,
\begin{eqnarray}
\begin{aligned}
|l_{\rho} \ l_{\lambda} \ L \ s\ j\ J\ M_{J}\rangle
&=\{[(|l_{\rho}\ m_{\rho} \rangle |l_{\lambda}\ m_{\lambda} \rangle)_{L}\times(|s_{1}\ m_{s_{1}} \rangle|s_{2}\ m_{s_{2}} \rangle)_{s}]_{j}\times|s_{3}\ m_{s_{3}} \rangle \}_{J M_{J}}.
\end{aligned}
\end{eqnarray}
$l_{\rho}$, $l_{\lambda}$, $L$, $s$, $j$, $J$ and $M_{J}$  are the quantum numbers. The total spin $s$ of ($bc$) and their orbital quantum number $l_{\rho}$ should meet the following condition: $(-1)^{s+l_{\rho}}=1$ for $\Xi_{bc}$($\Omega_{bc}$) and $(-1)^{s+l_{\rho}}=-1$ for $\Xi_{bc}^{'}$($\Omega_{bc}^{'}$), which guarantees the antisymmetry of the total wave function.

\section*{III. Numerical results and discussion}

\subsection*{3.1  $\rho$-mode  }

  As usual, $nL(J^{P})$ is used to describe a baryon state. For the excited states ($L\neq0$), there exist several $|l_{\rho} l_{\lambda}  L  s j J M_{J}\rangle$ states under the condition of $\textbf{L}=\textbf{\emph{l}}_{\rho}+\textbf{\emph{l}}_{\lambda}$. They may be divided into the following three modes: (1) The $\rho$-mode with $l_{\rho}\neq0$ and $l_{\lambda}=0$; (2) The $\lambda$-mode with $l_{\rho}=0$ and $l_{\lambda}\neq0$; (3) The $\lambda$-$\rho$ mixing mode with $l_{\rho}\neq0$ and $l_{\lambda}\neq0$.
 As an approximation, we take no account of the mixing of these modes, and note that the most likely mode to be observed experimentally should be that with lower energy. With this in mind, we need to analyze which mode has the lowest energy in our calculations.

 Considering the excitation energies of the $1D(\frac{3}{2}^{+}, \frac{5}{2}^{+})_{j=2}$ states for $\Xi_{QQ^{'}}^{'}$ and $\Xi_{QQ^{'}}$ as functions of $m_{2}$, we investigate the trend of these three modes in the heavy quark limit. Meanwhile, the dependence of excitation energies on $m_{2}$ of the $\rho$-mode is compared with that of the other two modes. Being $m_{c}=1.628$ GeV and $m_{b}=4.977$ GeV in the actual calculations below, we set $m_{1}=(m_{c}/m_{b})m_{2}$ here to keep them in proportion. From Fig.2, one can see that the excitation energies of the $\rho$-mode are significantly lower than those of the other two modes when $m_{2}$ increases from 2.0 GeV to 5.0 GeV.
  This suggests that the $\rho$-mode appears lower in energy than the other two modes in the heavy quark limit for bottom-charm baryons. Therefore, we only study the $\rho$-mode in this work.
  Additionally, the excitation energy differences between the $1D(\frac{3}{2}^{+}, \frac{5}{2}^{+})_{j=2}$ states in the $\rho$-mode get closer to each other with increasing $m_{2}$ as shown in Fig.2, which is consistent with the heavy quark spin symmetry~\cite{a0123}.

  \begin{figure}[htbp]
\begin{center}
\includegraphics[width=0.9\textwidth]{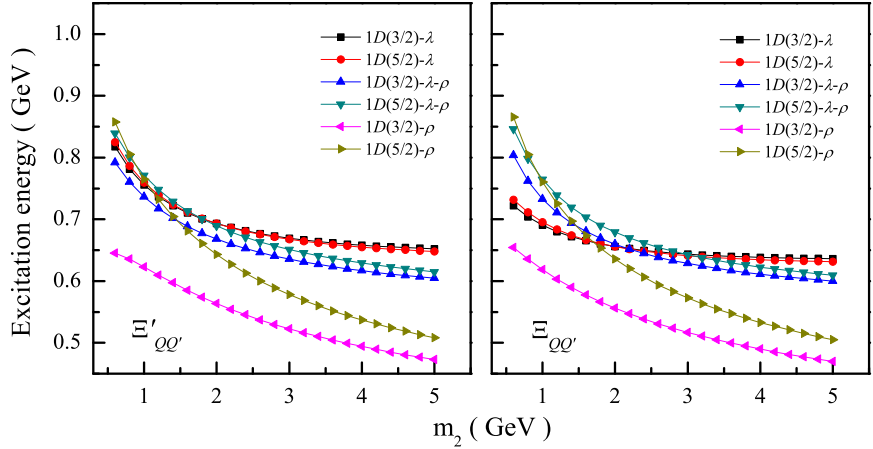}
\end{center}
\caption{(Color online) The dependence of excitation energies on $m_{2}$ for different modes of $\Xi_{QQ^{'}}^{'}$ and $\Xi_{QQ^{'}}$, where $m_{1}=(m_{c}/m_{b})m_{2}$ and $m_{3}=m_{u(d)}$. The excitation energies are measured from the ground states $1S(\frac{1}{2})^{+}_{j=1}$ of $\Xi_{QQ^{'}}^{'}$ and $1S(\frac{1}{2})^{+}_{j=0}$ of $\Xi_{QQ^{'}}$, respectively. }
\end{figure}

\subsection*{3.2 Mass spectra, r.m.s. radii and quark radial probability density distributions}

In the $\rho$-mode, the complete mass spectra of the $\Xi_{bc}^{'}$, $\Xi_{bc}$, $\Omega_{bc}^{'}$ and $\Omega_{bc}$ baryons with quantum numbers up to $n=4$ and $L=4$ are calculated. The corresponding r.m.s. radii and part of the quark radial probability density distributions are computed as well. The detailed results are presented in Tables I-IV and Figs.3-6.

 From Tables I-IV, one can find some general features as follows:
  (1) For the spin-doublet states with the same $j$, the $J=j+\frac{1}{2}$ state is higher in energy than the $J=j-\frac{1}{2}$ state.
  (2) For the same $L$, the mass splitting among different states becomes larger and larger with the increase of $j$. For example, Table I shows the mass differences (splittings) of the $1D$ doublets with $j=1,2,3$ are 21 MeV, 35 MeV and 50 MeV, respectively. So, the mass splitting of the bottom-charm baryons shows the same feature as those of the double-charm and -bottom baryons~\cite{a1055,a1056}. Moreover, by comparing the mass splittings of these three groups of the DHBs, one can find that this mass splitting is inversely proportional to the total mass of the heavy quark pair.
 (3) The mass difference between two adjacent radial excited states gradually decreases with increasing $n$. This is a general property for singly and doubly heavy baryons in our calculations.

The radial probability densities $\omega(r_{\rho})$ and $\omega(r_{\lambda})$ in a three-quark system can be defined below,
\begin{eqnarray}
\begin{aligned}
& \omega(r_{\rho})=\int |\Psi(\textbf{r}_{\rho},\textbf{r}_{\lambda})|^{2}\mathrm{d}\textbf{r}_{\lambda}\mathrm{d}\Omega_{\rho},\\
&  \omega(r_{\lambda})=\int |\Psi(\textbf{r}_{\rho},\textbf{r}_{\lambda})|^{2}\mathrm{d}\textbf{r}_{\rho}\mathrm{d}\Omega_{\lambda},
\end{aligned}
\end{eqnarray}
where $\Omega_{\rho}$ and $\Omega_{\lambda}$ stand for the solid angles spanned by vectors $\textbf{r}_{\rho}$ and $\textbf{r}_{\lambda}$, respectively. Through the analyses of the quark radial probability density distributions in Figs.3-6 and the r.m.s. radii in Tables I-IV, one can find some interesting properties.

(1) The $\langle r_{\rho}^{2}\rangle ^{1/2}$ value of a ground state is smaller than the corresponding $\langle r_{\lambda}^{2}\rangle ^{1/2}$ value. This means that the two heavy quarks are bonded tightly. By using the results of our previous papers~\cite{a1055,a1056} and examining the r.m.s. radii of these three groups of DHBs, one can find the $\langle r_{\rho}^{2}\rangle ^{1/2}$ value is also inversely proportional to the total mass of the heavy quark pair.

(2) Tables I-IV show that when $n$ is fixed, $\langle r_{\rho}^{2}\rangle ^{1/2}$ values become larger with increasing $L$.  Figs.3-6 also show a consistent change trend for the distribution of the $r_{\rho}^{2}\omega(r_{\rho})$ (solid lines),  especially in the case of $n$=1 or 3, where the peak of $r_{\rho}^{2}\omega(r_{\rho})$ is significantly shifted outward with increasing $L$.

(3)The curves in Fig.3 are very similar to those in Fig.4 and the values of the r.m.s. radii in Tables I and II are also very close to each other for the same $nL(J^{P})_{j}$ state. This suggests that the flavor symmetry has only a little effect on the shapes of the bottom-charm baryons and their mass values.

(4)Although the difference of the curves in Fig.3 (or 4) and Fig.5 (or 6) is very small, for the same quantum state, the apparent differences can be seen in the $\langle r_{\lambda}^{2}\rangle ^{1/2}$ values and the mass values for $\Xi_{bc}$ and $\Omega_{bc}$ as shown in Table I and III. This reflects the different contributions to these physical quantities from up (down) quark and strange quark in bottom-charm baryons.

\begin{table*}[htbp]
\begin{ruledtabular}\caption{The root mean square radii (fm) and mass spectra (MeV) of the $\Xi_{bc}^{'}$ family.}
\begin{tabular}{c c c c c | c c c c c}
$l_{\rho}$  $l_{\lambda}$ $L$ $s$ $j$  &$nL$($J^{P}$) & $\langle r_{\rho}^{2}\rangle^{1/2}$ & $\langle r_{\lambda}^{2}\rangle^{1/2}$ & mass & $l_{\rho}$  $l_{\lambda}$ $L$ $s$ $j$  &$nL$($J^{P}$) & $\langle r_{\rho}^{2}\rangle^{1/2}$ & $\langle r_{\lambda}^{2}\rangle^{1/2}$ & mass \\ \hline
\multirow{4}{*}{0 0 0 1 1 }
        & $1S$($\frac{1}{2}^{+}$)  & 0.379 & 0.469 & 6952 & \multirow{4}{*}{2 0 2 1 3} & $1D$($\frac{7}{2}^{+}$) & 0.749  & 0.566 & 7470 \\
        & $2S$($\frac{1}{2}^{+}$)  & 0.691 & 0.555 & 7346 & ~ & $2D$($\frac{7}{2}^{+}$)  & 0.952 & 0.602 & 7747 \\
        & $3S$($\frac{1}{2}^{+}$)  & 0.429 & 0.831 & 7455 & ~ & $3D$($\frac{7}{2}^{+}$)  & 0.792 & 0.925 & 7922  \\
        & $4S$($\frac{1}{2}^{+}$)  & 1.072 & 0.610 & 7673 & ~ & $4D$($\frac{7}{2}^{+}$)  & 1.521 & 0.685 & 8011    \\ \hline
\multirow{4}{*}{0 0 0 1 1}
        & $1S$($\frac{3}{2}^{+}$)  & 0.382 & 0.488& 6980 & \multirow{4}{*}{3 0 3 0 3} & $1F$($\frac{5}{2}^{-}$) & 0.888 & 0.549 & 7598 \\
        & $2S$($\frac{3}{2}^{+}$)  & 0.694 & 0.574& 7368 & ~ & $2F$($\frac{5}{2}^{-}$)  & 1.022 & 0.570 & 7862 \\
        & $3S$($\frac{3}{2}^{+}$)  & 0.432 & 0.841& 7470 & ~ & $3F$($\frac{5}{2}^{-}$)  & 0.941 & 0.922 & 8051  \\
        & $4S$($\frac{3}{2}^{+}$)  & 1.080 & 0.628& 7692 & ~ & $4F$($\frac{5}{2}^{-}$)  & 1.678 & 0.663 & 8109  \\ \hline
\multirow{4}{*}{1 0 1 0 1 }
        & $1P$($\frac{1}{2}^{-}$)  & 0.566 & 0.504& 7223 & \multirow{4}{*}{3 0 3 0 3} & $1F$($\frac{7}{2}^{-}$) & 0.899  & 0.592 & 7642\\
        & $2P$($\frac{1}{2}^{-}$)  & 0.838 & 0.561& 7538 & ~ & $2F$($\frac{7}{2}^{-}$)  & 1.035 & 0.611 & 7907 \\
        & $3P$($\frac{1}{2}^{-}$)  & 0.609 & 0.872& 7705 & ~ & $3F$($\frac{7}{2}^{-}$)  & 0.946 & 0.952 & 8082 \\
        & $4P$($\frac{1}{2}^{-}$)  & 1.321 & 0.637& 7840 & ~ & $4F$($\frac{7}{2}^{-}$)  & 1.668 & 0.700 & 8142 \\ \hline
\multirow{4}{*}{1 0 1 0 1 }
        & $1P$($\frac{3}{2}^{-}$)  & 0.571 & 0.523& 7247 & \multirow{4}{*}{4 0 4 1 3} & $1G$($\frac{5}{2}^{+}$) & 1.029 & 0.572 & 7759 \\
        & $2P$($\frac{3}{2}^{-}$)  & 0.841 & 0.579& 7559 & ~ & $2G$($\frac{5}{2}^{+}$)  & 1.097 & 0.579 & 8024 \\
        & $3P$($\frac{3}{2}^{-}$)  & 0.610 & 0.884& 7719 & ~ & $3G$($\frac{5}{2}^{+}$)  & 1.078 & 0.946 & 8199 \\
        & $4P$($\frac{3}{2}^{-}$)  & 1.326 & 0.654& 7856 & ~ & $4G$($\frac{5}{2}^{+}$)  & 1.806 & 0.675 & 8236  \\ \hline
\multirow{4}{*}{2 0 2 1 1}
        & $1D$($\frac{1}{2}^{+}$)  & 0.728 & 0.534& 7431& \multirow{4}{*}{4 0 4 1 3} & $1G$($\frac{7}{2}^{+}$) & 1.039 & 0.614 & 7800 \\
        & $2D$($\frac{1}{2}^{+}$)  & 0.940 & 0.572& 7708 & ~ & $2G$($\frac{7}{2}^{+}$)  & 1.121 & 0.621 & 8067 \\
        & $3D$($\frac{1}{2}^{+}$)  & 0.777 & 0.902& 7896 & ~ & $3G$($\frac{7}{2}^{+}$)  & 1.085 & 0.975 & 8228 \\
        & $4D$($\frac{1}{2}^{+}$)  & 1.522 & 0.660& 7986 & ~ & $4G$($\frac{7}{2}^{+}$)  & 1.786 & 0.710 & 8267 \\ \hline
\multirow{4}{*}{2 0 2 1 1}
        & $1D$($\frac{3}{2}^{+}$)  & 0.733 & 0.552& 7452 & \multirow{4}{*}{4 0 4 1 4} & $1G$($\frac{7}{2}^{+}$) & 1.030 & 0.566 & 7751 \\
        & $2D$($\frac{3}{2}^{+}$)  & 0.943 & 0.590& 7728 & ~ & $2G$($\frac{7}{2}^{+}$)  & 1.097 & 0.572 & 8017 \\
        & $3D$($\frac{3}{2}^{+}$)  & 0.778 & 0.915& 7909 & ~ & $3G$($\frac{7}{2}^{+}$)  & 1.080 & 0.942 & 8193 \\
        & $4D$($\frac{3}{2}^{+}$)  & 1.521 & 0.676& 8001 & ~ & $4G$($\frac{7}{2}^{+}$)  & 1.806 & 0.668 & 8229  \\ \hline
\multirow{4}{*}{2 0 2 1 2}
        & $1D$($\frac{3}{2}^{+}$)  & 0.731 & 0.528& 7425 & \multirow{4}{*}{4 0 4 1 4 } & $1G$($\frac{9}{2}^{+}$) & 1.043  & 0.620 & 7803 \\
        & $2D$($\frac{3}{2}^{+}$) & 0.941 & 0.566 & 7703 & ~ & $2G$($\frac{9}{2}^{+}$)  & 1.129 & 0.628 & 8073 \\
        & $3D$($\frac{3}{2}^{+}$)  & 0.781 & 0.899& 7892 & ~ & $3G$($\frac{9}{2}^{+}$)  & 1.088 & 0.980 & 8230 \\
        & $4D$($\frac{3}{2}^{+}$)  & 1.522 & 0.654& 7981 & ~ & $4G$($\frac{9}{2}^{+}$)  & 1.780 & 0.714 & 8270    \\ \hline
\multirow{4}{*}{2 0 2 1 2 }
        & $1D$($\frac{5}{2}^{+}$)  & 0.740 & 0.559& 7460 & \multirow{4}{*}{4 0 4 1 5} & $1G$($\frac{9}{2}^{+}$) & 1.032 & 0.560 & 7743 \\
        & $2D$($\frac{5}{2}^{+}$)  & 0.946 & 0.596& 7736 & ~ & $2G$($\frac{9}{2}^{+}$)  & 1.100 & 0.566 & 8011 \\
        & $3D$($\frac{5}{2}^{+}$) & 0.784 & 0.920 & 7915 & ~ & $3G$($\frac{9}{2}^{+}$)  & 1.083 & 0.939 & 8186 \\
        & $4D$($\frac{5}{2}^{+}$)  & 1.521 & 0.681& 8006 & ~ & $4G$($\frac{9}{2}^{+}$)  & 1.804 & 0.662 & 8223 \\ \hline
\multirow{4}{*}{2 0 2 1 3 }
        & $1D$($\frac{5}{2}^{+}$)  & 0.737 & 0.523& 7420 & \multirow{4}{*}{4 0 4 1 5} & $1G$($\frac{11}{2}^{+}$) & 1.047 & 0.626 & 7807\\
        & $2D$($\frac{5}{2}^{+}$)  & 0.943 & 0.561& 7700 & ~ & $2G$($\frac{11}{2}^{+}$)  & 1.141 & 0.634 & 8079 \\
        & $3D$($\frac{5}{2}^{+}$)  & 0.788 & 0.896& 7890 & ~ & $3G$($\frac{11}{2}^{+}$)  & 1.093 & 0.984 & 8232 \\
        & $4D$($\frac{5}{2}^{+}$)  & 1.524 & 0.648& 7976 & ~ & $4G$($\frac{11}{2}^{+}$)  & 1.769 & 0.717 & 8273 \\
\end{tabular}
\end{ruledtabular}
\end{table*}
\begin{table*}[htbp]
\begin{ruledtabular}\caption{The root mean square radii (fm) and mass spectra (MeV) of the $\Xi_{bc}$ family.}
\begin{tabular}{c c c c c | c c c c c}
$l_{\rho}$  $l_{\lambda}$ $L$ $s$ $j$  &$nL$($J^{P}$) & $\langle r_{\rho}^{2}\rangle^{1/2}$ & $\langle r_{\lambda}^{2}\rangle^{1/2}$ & mass & $l_{\rho}$  $l_{\lambda}$ $L$ $s$ $j$  &$nL$($J^{P}$) & $\langle r_{\rho}^{2}\rangle^{1/2}$ & $\langle r_{\lambda}^{2}\rangle^{1/2}$ & mass \\ \hline
\multirow{4}{*}{0 0 0 0 0 }
        & $1S$($\frac{1}{2}^{+}$)  & 0.370 & 0.479 & 6955 & \multirow{4}{*}{3 0 3 1 2} & $1F$($\frac{3}{2}^{-}$) & 0.885  & 0.555 & 7605 \\
        & $2S$($\frac{1}{2}^{+}$)  & 0.683 & 0.569 & 7351 & ~ & $2F$($\frac{3}{2}^{-}$)  & 1.021 & 0.575 & 7868 \\
        & $3S$($\frac{1}{2}^{+}$)  & 0.422 & 0.834 & 7451 & ~ & $3F$($\frac{3}{2}^{-}$)  & 0.937 & 0.926 & 8056  \\
        & $4S$($\frac{1}{2}^{+}$)  & 1.060 & 0.621 & 7677 & ~ & $4F$($\frac{3}{2}^{-}$)  & 1.679 & 0.670 & 8115    \\ \hline
\multirow{4}{*}{1 0 1 1 0 }
        & $1P$($\frac{1}{2}^{-}$)  & 0.559 & 0.514& 7231 & \multirow{4}{*}{3 0 3 1 2} & $1F$($\frac{5}{2}^{-}$) & 0.893  & 0.585 & 7637\\
        & $2P$($\frac{1}{2}^{-}$)  & 0.835 & 0.572& 7545 & ~ & $2F$($\frac{5}{2}^{-}$)  & 1.030 & 0.605 & 7900 \\
        & $3P$($\frac{1}{2}^{-}$)  & 0.599 & 0.878& 7707 & ~ & $3F$($\frac{5}{2}^{-}$)  & 0.940 & 0.947 & 8078 \\
        & $4P$($\frac{1}{2}^{-}$)  & 1.315 & 0.648& 7847 & ~ & $4F$($\frac{5}{2}^{-}$)  & 1.672 & 0.696 & 8138 \\ \hline
\multirow{4}{*}{1 0 1 1 1 }
        & $1P$($\frac{1}{2}^{-}$)  & 0.562 & 0.503& 7220 & \multirow{4}{*}{3 0 3 1 3} & $1F$($\frac{5}{2}^{-}$) & 0.886  & 0.549 & 7598\\
        & $2P$($\frac{1}{2}^{-}$)  & 0.836 & 0.561& 7535 & ~ & $2F$($\frac{5}{2}^{-}$)  & 1.021 & 0.570 & 7862 \\
        & $3P$($\frac{1}{2}^{-}$)  & 0.604 & 0.871& 7702 & ~ & $3F$($\frac{5}{2}^{-}$)  & 0.940 & 0.922 & 8051 \\
        & $4P$($\frac{1}{2}^{-}$)  & 1.317 & 0.637& 7838 & ~ & $4F$($\frac{5}{2}^{-}$)  & 1.679 & 0.664 & 8109 \\ \hline
\multirow{4}{*}{1 0 1 1 1 }
        & $1P$($\frac{3}{2}^{-}$)  & 0.566 & 0.522& 7243 & \multirow{4}{*}{3 0 3 1 3} & $1F$($\frac{7}{2}^{-}$) & 0.898 & 0.591 & 7642 \\
        & $2P$($\frac{3}{2}^{-}$)  & 0.839 & 0.578& 7556 & ~ & $2F$($\frac{7}{2}^{-}$)  & 1.034 & 0.611 & 7907 \\
        & $3P$($\frac{3}{2}^{-}$)  & 0.605 & 0.883& 7716 & ~ & $3F$($\frac{7}{2}^{-}$)  & 0.944 & 0.951 & 8082 \\
        & $4P$($\frac{3}{2}^{-}$)  & 1.322 & 0.654& 7855 & ~ & $4F$($\frac{7}{2}^{-}$)  & 1.668 & 0.700 & 8142  \\ \hline
\multirow{4}{*}{1 0 1 1 2}
        & $1P$($\frac{3}{2}^{-}$)  & 0.572 & 0.499& 7221 & \multirow{4}{*}{3 0 3 1 4} & $1F$($\frac{7}{2}^{-}$) & 0.890 & 0.543 & 7591 \\
        & $2P$($\frac{3}{2}^{-}$)  & 0.841 & 0.556& 7536 & ~ & $2F$($\frac{7}{2}^{-}$)  & 1.023 & 0.564 & 7857 \\
        & $3P$($\frac{3}{2}^{-}$)  & 0.616 & 0.869& 7706 & ~ & $3F$($\frac{7}{2}^{-}$)  & 0.944 & 0.919 & 8047 \\
        & $4P$($\frac{3}{2}^{-}$)  & 1.327 & 0.632& 7837 & ~ & $4F$($\frac{7}{2}^{-}$)  & 1.678 & 0.657 & 8103 \\ \hline
\multirow{4}{*}{1 0 1 1 2}
        & $1P$($\frac{5}{2}^{-}$)  & 0.580 & 0.530& 7261 & \multirow{4}{*}{3 0 3 1 4} & $1F$($\frac{9}{2}^{-}$) & 0.905 & 0.598 & 7648 \\
        & $2P$($\frac{5}{2}^{-}$)  & 0.846 & 0.585& 7571 & ~ & $2F$($\frac{9}{2}^{-}$)  & 1.041 & 0.617 & 7914 \\
        & $3P$($\frac{5}{2}^{-}$)  & 0.618 & 0.890& 7730 & ~ & $3F$($\frac{9}{2}^{-}$)  & 0.951 & 0.957 & 8086 \\
        & $4P$($\frac{5}{2}^{-}$)  & 1.334 & 0.660& 7865 & ~ & $4F$($\frac{9}{2}^{-}$)  & 1.663 & 0.704 & 8145  \\ \hline
\multirow{4}{*}{2 0 2 0 2}
        & $1D$($\frac{3}{2}^{+}$)  & 0.733 & 0.529& 7426 & \multirow{4}{*}{4 0 4 0 4 } & $1G$($\frac{7}{2}^{+}$) & 1.030  & 0.566 & 7751 \\
        & $2D$($\frac{3}{2}^{+}$) & 0.942 & 0.567 & 7704 & ~ & $2G$($\frac{7}{2}^{+}$)  & 1.099 & 0.573 & 8017 \\
        & $3D$($\frac{3}{2}^{+}$)  & 0.783 & 0.899& 7893 & ~ & $3G$($\frac{7}{2}^{+}$)  & 1.081 & 0.942 & 8192 \\
        & $4D$($\frac{3}{2}^{+}$)  & 1.523 & 0.654& 7981 & ~ & $4G$($\frac{7}{2}^{+}$)  & 1.805 & 0.668 & 8229    \\ \hline
\multirow{4}{*}{2 0 2 0 2 }
        & $1D$($\frac{5}{2}^{+}$)  & 0.742 & 0.559& 7461 & \multirow{4}{*}{4 0 4 0 4} & $1G$($\frac{9}{2}^{+}$) & 1.043 & 0.620 & 7803 \\
        & $2D$($\frac{5}{2}^{+}$)  & 0.948 & 0.596& 7737 & ~ & $2G$($\frac{9}{2}^{+}$)  & 1.131 & 0.628 & 8073 \\
        & $3D$($\frac{5}{2}^{+}$) & 0.786 & 0.920 & 7916 & ~ & $3G$($\frac{9}{2}^{+}$)  & 1.089 & 0.980 & 8230 \\
        & $4D$($\frac{5}{2}^{+}$)  & 1.521 & 0.681& 8006 & ~ & $4G$($\frac{9}{2}^{+}$)  & 1.778 & 0.714 & 8270 \\
\end{tabular}
\end{ruledtabular}
\end{table*}
\begin{table*}[htbp]
\begin{ruledtabular}\caption{The root mean square radii (fm) and mass spectra (MeV) of the $\Omega_{bc}^{'}$ family.}
\begin{tabular}{c c c c c | c c c c c}
$l_{\rho}$  $l_{\lambda}$ $L$ $s$ $j$  &$nL$($J^{P}$) & $\langle r_{\rho}^{2}\rangle^{1/2}$ & $\langle r_{\lambda}^{2}\rangle^{1/2}$ & mass & $l_{\rho}$  $l_{\lambda}$ $L$ $s$ $j$  &$nL$($J^{P}$) & $\langle r_{\rho}^{2}\rangle^{1/2}$ & $\langle r_{\lambda}^{2}\rangle^{1/2}$ & mass \\ \hline
\multirow{4}{*}{0 0 0 1 1 }
        & $1S$($\frac{1}{2}^{+}$)  & 0.371 & 0.429 & 7053 & \multirow{4}{*}{2 0 2 1 3} & $1D$($\frac{7}{2}^{+}$) & 0.738  & 0.522 & 7578 \\
        & $2S$($\frac{1}{2}^{+}$)  & 0.675 & 0.526 & 7453 & ~ & $2D$($\frac{7}{2}^{+}$)  & 0.942 & 0.558 & 7856 \\
        & $3S$($\frac{1}{2}^{+}$)  & 0.438 & 0.771 & 7554 & ~ & $3D$($\frac{7}{2}^{+}$)  & 0.787 & 0.872 & 8022  \\
        & $4S$($\frac{1}{2}^{+}$)  & 1.039 & 0.581 & 7786 & ~ & $4D$($\frac{7}{2}^{+}$)  & 1.521 & 0.639 & 8131    \\ \hline
\multirow{4}{*}{0 0 0 1 1}
        & $1S$($\frac{3}{2}^{+}$)  & 0.375 & 0.446& 7079 & \multirow{4}{*}{3 0 3 0 3} & $1F$($\frac{5}{2}^{-}$) & 0.877 & 0.510 & 7716 \\
        & $2S$($\frac{3}{2}^{+}$)  & 0.677 & 0.542& 7474 & ~ & $2F$($\frac{5}{2}^{-}$)  & 1.011 & 0.532 & 7979 \\
        & $3S$($\frac{3}{2}^{+}$)  & 0.441 & 0.782& 7568 & ~ & $3F$($\frac{5}{2}^{-}$)  & 0.933 & 0.871 & 8155  \\
        & $4S$($\frac{3}{2}^{+}$)  & 1.048 & 0.596& 7803 & ~ & $4F$($\frac{5}{2}^{-}$)  & 1.685 & 0.621 & 8236  \\ \hline
\multirow{4}{*}{1 0 1 0 1 }
        & $1P$($\frac{1}{2}^{-}$)  & 0.557 & 0.465& 7331 & \multirow{4}{*}{3 0 3 0 3} & $1F$($\frac{7}{2}^{-}$) & 0.889  & 0.546 & 7754\\
        & $2P$($\frac{1}{2}^{-}$)  & 0.829 & 0.524& 7649 & ~ & $2F$($\frac{7}{2}^{-}$)  & 1.021 & 0.567 & 8018 \\
        & $3P$($\frac{1}{2}^{-}$)  & 0.607 & 0.818& 7806 & ~ & $3F$($\frac{7}{2}^{-}$)  & 0.938 & 0.900 & 8183 \\
        & $4P$($\frac{1}{2}^{-}$)  & 1.301 & 0.599& 7958 & ~ & $4F$($\frac{7}{2}^{-}$)  & 1.678 & 0.652 & 8263 \\ \hline
\multirow{4}{*}{1 0 1 0 1 }
        & $1P$($\frac{3}{2}^{-}$)  & 0.561 & 0.481& 7353 & \multirow{4}{*}{4 0 4 1 3} & $1G$($\frac{5}{2}^{+}$) & 1.019 & 0.532 & 7879 \\
        & $2P$($\frac{3}{2}^{-}$)  & 0.832 & 0.539& 7668 & ~ & $2G$($\frac{5}{2}^{+}$)  & 1.078 & 0.539 & 8141 \\
        & $3P$($\frac{3}{2}^{-}$)  & 0.608 & 0.830& 7819 & ~ & $3G$($\frac{5}{2}^{+}$)  & 1.069 & 0.896 & 8303 \\
        & $4P$($\frac{3}{2}^{-}$)  & 1.308 & 0.613& 7973 & ~ & $4G$($\frac{5}{2}^{+}$)  & 1.822 & 0.631 & 8364  \\ \hline
\multirow{4}{*}{2 0 2 1 1}
        & $1D$($\frac{1}{2}^{+}$)  & 0.718 & 0.494& 7543 & \multirow{4}{*}{4 0 4 1 3} & $1G$($\frac{7}{2}^{+}$) & 1.029 & 0.568 & 7914 \\
        & $2D$($\frac{1}{2}^{+}$)  & 0.932 & 0.534& 7822 & ~ & $2G$($\frac{7}{2}^{+}$)  & 1.095 & 0.575 & 8178 \\
        & $3D$($\frac{1}{2}^{+}$)  & 0.771 & 0.849& 7997 & ~ & $3G$($\frac{7}{2}^{+}$)  & 1.074 & 0.925 & 8330 \\
        & $4D$($\frac{1}{2}^{+}$)  & 1.516 & 0.618& 8109 & ~ & $4G$($\frac{7}{2}^{+}$)  & 1.808 & 0.661 & 8390 \\ \hline
\multirow{4}{*}{2 0 2 1 1}
        & $1D$($\frac{3}{2}^{+}$)  & 0.723 & 0.510& 7562 & \multirow{4}{*}{4 0 4 1 4} & $1G$($\frac{7}{2}^{+}$) & 1.020 & 0.526 & 7872 \\
        & $2D$($\frac{3}{2}^{+}$)  & 0.935 & 0.548& 7839 & ~ & $2G$($\frac{7}{2}^{+}$)  & 1.079 & 0.534 & 8135 \\
        & $3D$($\frac{3}{2}^{+}$)  & 0.773 & 0.862& 8010 & ~ & $3G$($\frac{7}{2}^{+}$)  & 1.070 & 0.892 & 8298 \\
        & $4D$($\frac{3}{2}^{+}$)  & 1.518 & 0.632& 8122 & ~ & $4G$($\frac{7}{2}^{+}$)  & 1.822 & 0.626 & 8359  \\ \hline
\multirow{4}{*}{2 0 2 1 2}
        & $1D$($\frac{3}{2}^{+}$)  & 0.720 & 0.489& 7538 & \multirow{4}{*}{4 0 4 1 4 } & $1G$($\frac{9}{2}^{+}$) & 1.033  & 0.572 & 7917 \\
        & $2D$($\frac{3}{2}^{+}$) & 0.933 & 0.529 & 7818 & ~ & $2G$($\frac{9}{2}^{+}$)  & 1.102 & 0.580 & 8183 \\
        & $3D$($\frac{3}{2}^{+}$)  & 0.775 & 0.846& 7994 & ~ & $3G$($\frac{9}{2}^{+}$)  & 1.078 & 0.929 & 8332 \\
        & $4D$($\frac{3}{2}^{+}$)  & 1.518 & 0.613& 8105 & ~ & $4G$($\frac{9}{2}^{+}$)  & 1.803 & 0.665 & 8392    \\ \hline
\multirow{4}{*}{2 0 2 1 2 }
        & $1D$($\frac{5}{2}^{+}$)  & 0.729 & 0.516& 7569 & \multirow{4}{*}{4 0 4 1 5} & $1G$($\frac{9}{2}^{+}$) & 1.023 & 0.521 & 7865 \\
        & $2D$($\frac{5}{2}^{+}$)  & 0.937 & 0.553& 7847 & ~ & $2G$($\frac{9}{2}^{+}$)  & 1.080 & 0.529 & 8130 \\
        & $3D$($\frac{5}{2}^{+}$) & 0.778 & 0.866 & 8015 & ~ & $3G$($\frac{9}{2}^{+}$)  & 1.073 & 0.889 & 8292 \\
        & $4D$($\frac{5}{2}^{+}$)  & 1.519 & 0.636& 8126 & ~ & $4G$($\frac{9}{2}^{+}$)  & 1.820 & 0.620 & 8353 \\ \hline
\multirow{4}{*}{2 0 2 1 3 }
        & $1D$($\frac{5}{2}^{+}$)  & 0.726 & 0.485& 7535 & \multirow{4}{*}{4 0 4 1 5} & $1G$($\frac{11}{2}^{+}$) & 1.038 & 0.577 & 7920\\
        & $2D$($\frac{5}{2}^{+}$)  & 0.935 & 0.524& 7816 & ~ & $2G$($\frac{11}{2}^{+}$)  & 1.111 & 0.586 & 8189 \\
        & $3D$($\frac{5}{2}^{+}$)  & 0.783 & 0.844& 7993 & ~ & $3G$($\frac{11}{2}^{+}$)  & 1.082 & 0.933 & 8333 \\
        & $4D$($\frac{5}{2}^{+}$)  & 1.520 & 0.608& 8101 & ~ & $4G$($\frac{11}{2}^{+}$)  & 1.795 & 0.667 & 8394 \\
\end{tabular}
\end{ruledtabular}
\end{table*}
\begin{table*}[htbp]
\begin{ruledtabular}\caption{The root mean square radii (fm) and mass spectra (MeV) of the $\Omega_{bc}$ family.}
\begin{tabular}{c c c c c | c c c c c}
$l_{\rho}$  $l_{\lambda}$ $L$ $s$ $j$  &$nL$($J^{P}$) & $\langle r_{\rho}^{2}\rangle^{1/2}$ & $\langle r_{\lambda}^{2}\rangle^{1/2}$ & mass & $l_{\rho}$  $l_{\lambda}$ $L$ $s$ $j$  &$nL$($J^{P}$) & $\langle r_{\rho}^{2}\rangle^{1/2}$ & $\langle r_{\lambda}^{2}\rangle^{1/2}$ & mass \\ \hline
\multirow{4}{*}{0 0 0 0 0 }
        & $1S$($\frac{1}{2}^{+}$)  & 0.363 & 0.438 & 7055 & \multirow{4}{*}{3 0 3 1 2} & $1F$($\frac{3}{2}^{-}$) & 0.874  & 0.515 & 7722 \\
        & $2S$($\frac{1}{2}^{+}$)  & 0.666 & 0.538 & 7457 & ~ & $2F$($\frac{3}{2}^{-}$)  & 1.010 & 0.537 & 7984 \\
        & $3S$($\frac{1}{2}^{+}$)  & 0.433 & 0.774 & 7550 & ~ & $3F$($\frac{3}{2}^{-}$)  & 0.929 & 0.875 & 8160  \\
        & $4S$($\frac{1}{2}^{+}$)  & 1.026 & 0.591 & 7788 & ~ & $4F$($\frac{3}{2}^{-}$)  & 1.686 & 0.627 & 8241    \\ \hline
\multirow{4}{*}{1 0 1 1 0 }
        & $1P$($\frac{1}{2}^{-}$)  & 0.549 & 0.474& 7337 & \multirow{4}{*}{3 0 3 1 2} & $1F$($\frac{5}{2}^{-}$) & 0.882  & 0.541 & 7749\\
        & $2P$($\frac{1}{2}^{-}$)  & 0.826 & 0.534& 7655 & ~ & $2F$($\frac{5}{2}^{-}$)  & 1.016 & 0.562 & 8011 \\
        & $3P$($\frac{1}{2}^{-}$)  & 0.597 & 0.824& 7807 & ~ & $3F$($\frac{5}{2}^{-}$)  & 0.933 & 0.895 & 8180 \\
        & $4P$($\frac{1}{2}^{-}$)  & 1.293 & 0.608& 7964 & ~ & $4F$($\frac{5}{2}^{-}$)  & 1.681 & 0.648 & 8261 \\ \hline
\multirow{4}{*}{1 0 1 1 1 }
        & $1P$($\frac{1}{2}^{-}$)  & 0.552 & 0.464& 7327 & \multirow{4}{*}{3 0 3 1 3} & $1F$($\frac{5}{2}^{-}$) & 0.876  & 0.510 & 7715\\
        & $2P$($\frac{1}{2}^{-}$)  & 0.827 & 0.524& 7647 & ~ & $2F$($\frac{5}{2}^{-}$)  & 1.010 & 0.532 & 7978 \\
        & $3P$($\frac{1}{2}^{-}$)  & 0.602 & 0.817& 7803 & ~ & $3F$($\frac{5}{2}^{-}$)  & 0.932 & 0.871 & 8155 \\
        & $4P$($\frac{1}{2}^{-}$)  & 1.296 & 0.599& 7957 & ~ & $4F$($\frac{5}{2}^{-}$)  & 1.686 & 0.621 & 8236 \\ \hline
\multirow{4}{*}{1 0 1 1 1 }
        & $1P$($\frac{3}{2}^{-}$)  & 0.557 & 0.480& 7349 & \multirow{4}{*}{3 0 3 1 3} & $1F$($\frac{7}{2}^{-}$) & 0.887 & 0.546 & 7754 \\
        & $2P$($\frac{3}{2}^{-}$)  & 0.830 & 0.539& 7665 & ~ & $2F$($\frac{7}{2}^{-}$)  & 1.020 & 0.567 & 8017 \\
        & $3P$($\frac{3}{2}^{-}$)  & 0.603 & 0.829& 7816 & ~ & $3F$($\frac{7}{2}^{-}$)  & 0.937 & 0.900 & 8183 \\
        & $4P$($\frac{3}{2}^{-}$)  & 1.303 & 0.613& 7971 & ~ & $4F$($\frac{7}{2}^{-}$)  & 1.678 & 0.652 & 8263  \\ \hline
\multirow{4}{*}{1 0 1 1 2}
        & $1P$($\frac{3}{2}^{-}$)  & 0.563 & 0.460& 7330 & \multirow{4}{*}{3 0 3 1 4} & $1F$($\frac{7}{2}^{-}$) & 0.880 & 0.505 & 7710 \\
        & $2P$($\frac{3}{2}^{-}$)  & 0.832 & 0.519& 7648 & ~ & $2F$($\frac{7}{2}^{-}$)  & 1.012 & 0.526 & 7974 \\
        & $3P$($\frac{3}{2}^{-}$)  & 0.614 & 0.816& 7807 & ~ & $3F$($\frac{7}{2}^{-}$)  & 0.936 & 0.868 & 8151 \\
        & $4P$($\frac{3}{2}^{-}$)  & 1.309 & 0.595& 7957 & ~ & $4F$($\frac{7}{2}^{-}$)  & 1.685 & 0.616 & 8231 \\ \hline
\multirow{4}{*}{1 0 1 1 2}
        & $1P$($\frac{5}{2}^{-}$)  & 0.571 & 0.488& 7366 & \multirow{4}{*}{3 0 3 1 4} & $1F$($\frac{9}{2}^{-}$) & 0.894 & 0.552 & 7759 \\
        & $2P$($\frac{5}{2}^{-}$)  & 0.837 & 0.544& 7679 & ~ & $2F$($\frac{9}{2}^{-}$)  & 1.025 & 0.572 & 8024 \\
        & $3P$($\frac{5}{2}^{-}$)  & 0.616 & 0.835& 7829 & ~ & $3F$($\frac{9}{2}^{-}$)  & 0.943 & 0.904 & 8186 \\
        & $4P$($\frac{5}{2}^{-}$)  & 1.318 & 0.618& 7981 & ~ & $4F$($\frac{9}{2}^{-}$)  & 1.675 & 0.655 & 8266  \\ \hline
\multirow{4}{*}{2 0 2 0 2}
        & $1D$($\frac{3}{2}^{+}$)  & 0.723 & 0.489& 7539 & \multirow{4}{*}{4 0 4 0 4 } & $1G$($\frac{7}{2}^{+}$) & 1.021  & 0.526 & 7872 \\
        & $2D$($\frac{3}{2}^{+}$) & 0.934 & 0.529 & 7819 & ~ & $2G$($\frac{7}{2}^{+}$)  & 1.079 & 0.534 & 8135 \\
        & $3D$($\frac{3}{2}^{+}$)  & 0.777 & 0.847& 7995 & ~ & $3G$($\frac{7}{2}^{+}$)  & 1.071 & 0.892 & 8297 \\
        & $4D$($\frac{3}{2}^{+}$)  & 1.519 & 0.613& 8105 & ~ & $4G$($\frac{7}{2}^{+}$)  & 1.821 & 0.626 & 8358    \\ \hline
\multirow{4}{*}{2 0 2 0 2 }
        & $1D$($\frac{5}{2}^{+}$)  & 0.731 & 0.516& 7570 & \multirow{4}{*}{4 0 4 0 4} & $1G$($\frac{9}{2}^{+}$) & 1.034 & 0.573 & 7917 \\
        & $2D$($\frac{5}{2}^{+}$)  & 0.938 & 0.554& 7848 & ~ & $2G$($\frac{9}{2}^{+}$)  & 1.103 & 0.581 & 8183 \\
        & $3D$($\frac{5}{2}^{+}$) & 0.781 & 0.867 & 8016 & ~ & $3G$($\frac{9}{2}^{+}$)  & 1.078 & 0.929 & 8331 \\
        & $4D$($\frac{5}{2}^{+}$)  & 1.519 & 0.636& 8126 & ~ & $4G$($\frac{9}{2}^{+}$)  & 1.802 & 0.664 & 8392 \\
\end{tabular}
\end{ruledtabular}
\end{table*}
\begin{figure}[htbp]
\begin{center}
\includegraphics[width=0.9\textwidth]{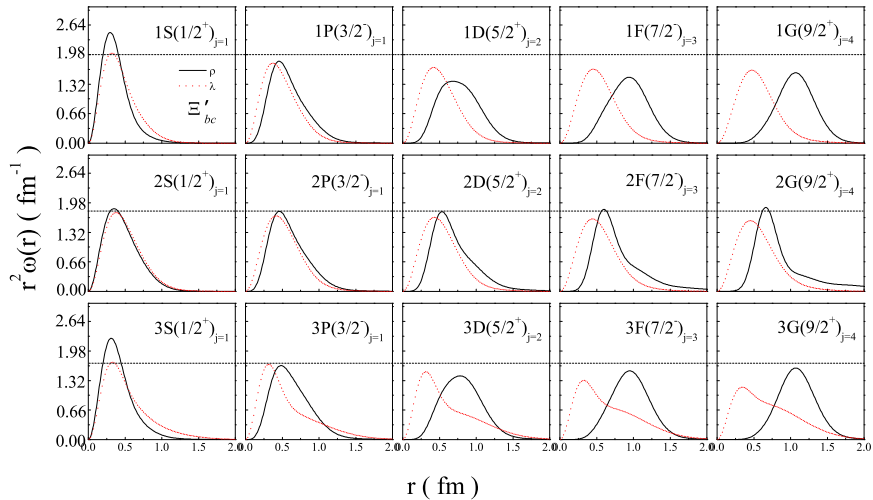}
\end{center}
\caption{(Color online)Quark radial probability density distributions for some $nL$ states in the $\Xi_{bc}^{'}$ family. The solid line denotes the probability density with $r_{\rho}$, and the dash line the one with $r_{\lambda}$.}
\end{figure}
\begin{figure}[htbp]
\begin{center}
\includegraphics[width=0.9\textwidth]{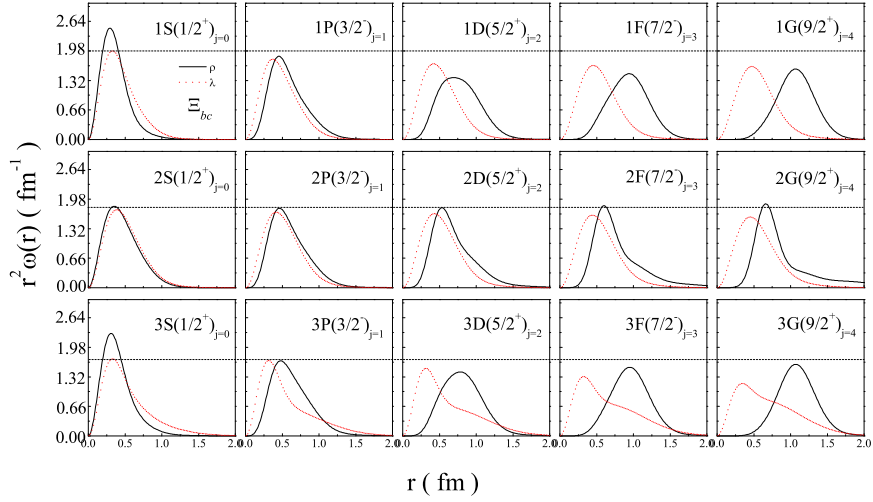}
\end{center}
\caption{(Color online)Same as Fig.3, but for the $\Xi_{bc}$ family.}
\end{figure}
\begin{figure}[htbp]
\begin{center}
\includegraphics[width=0.9\textwidth]{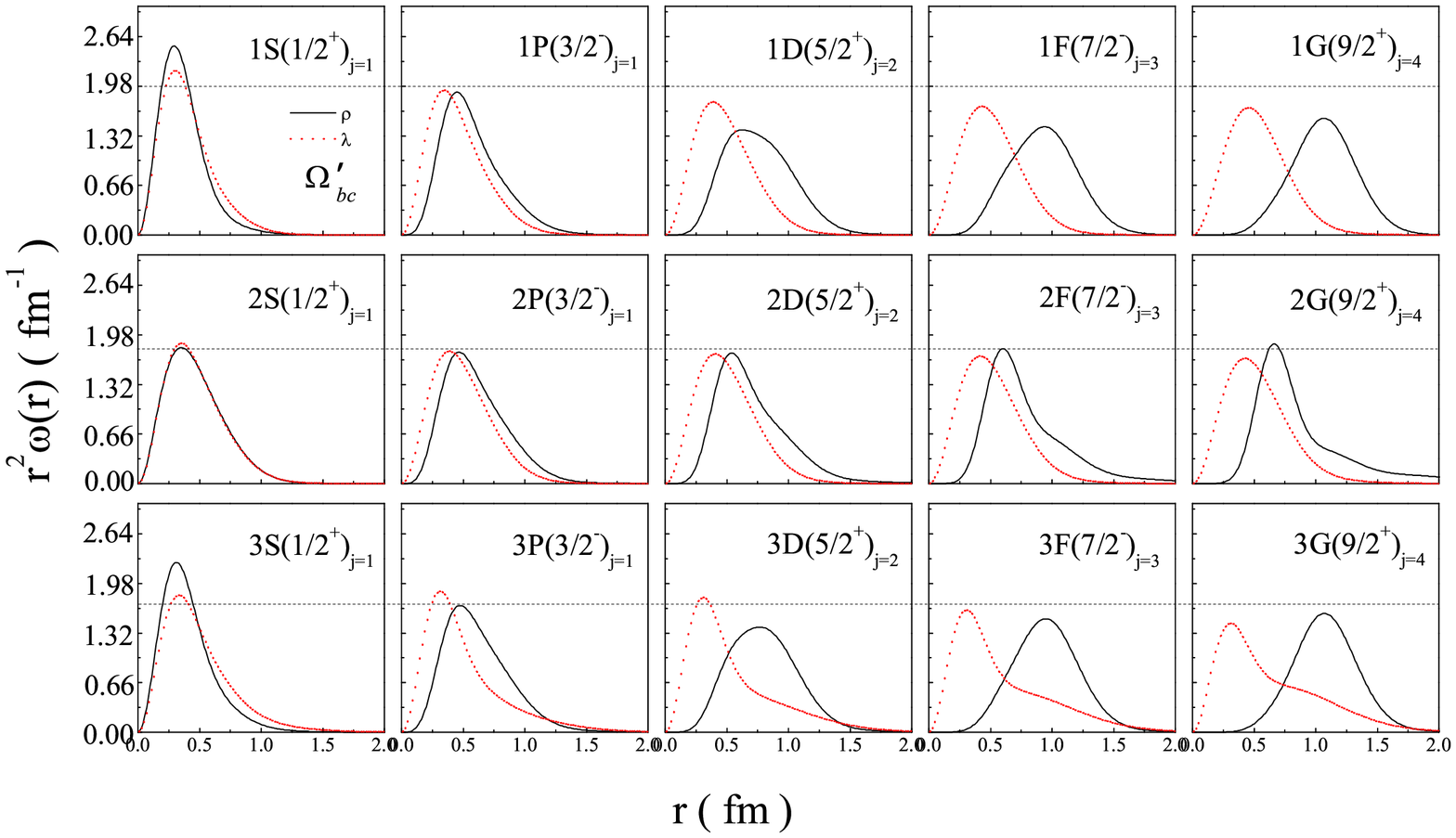}
\end{center}
\caption{(Color online)Same as Fig.3, but for the $\Omega_{bc}^{'}$ family.}
\end{figure}
\begin{figure}[htbp]
\begin{center}
\includegraphics[width=0.9\textwidth]{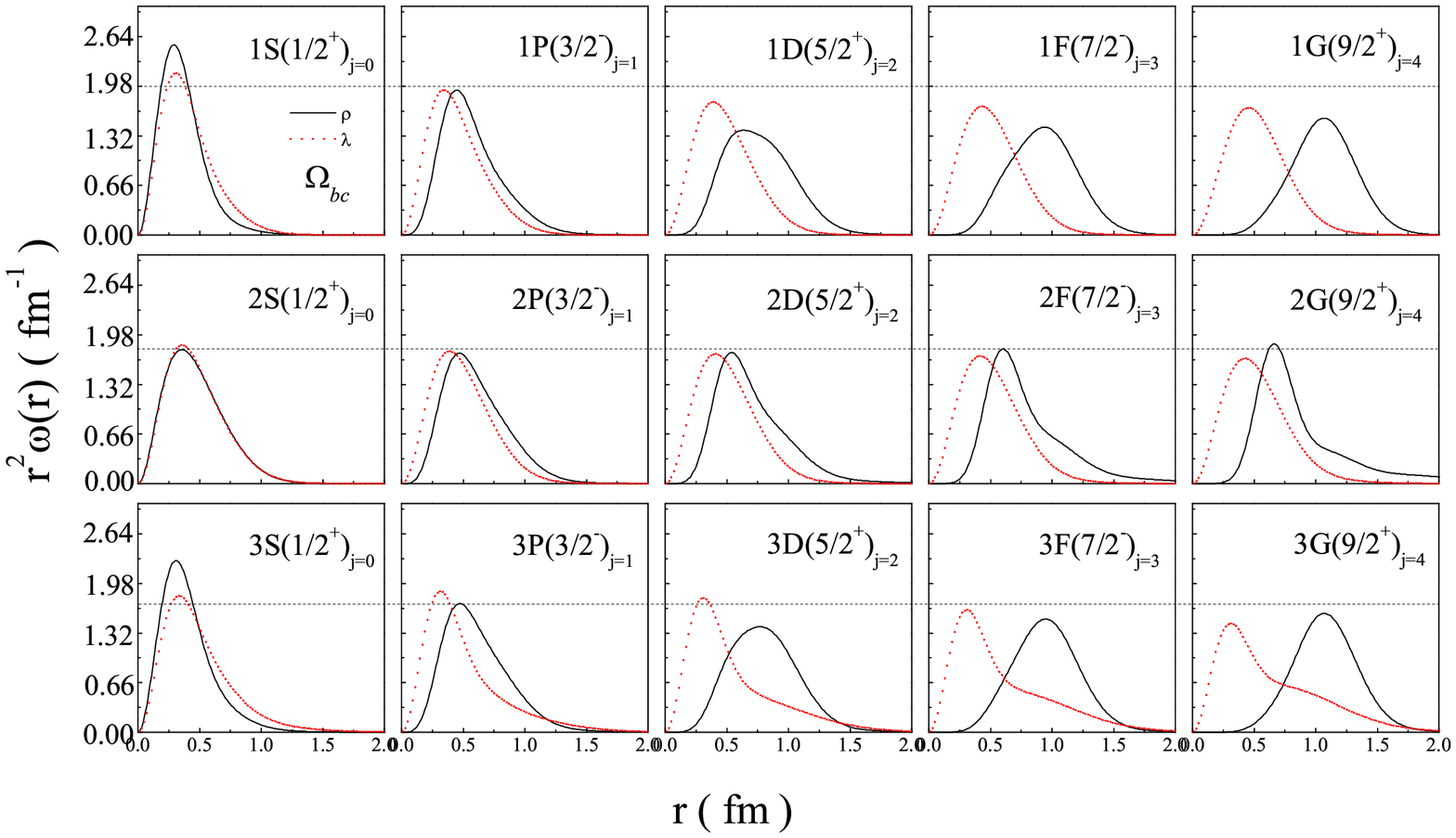}
\end{center}
\caption{(Color online)Same as Fig.3, but for the $\Omega_{bc}$ family.}
\end{figure}

\subsection*{3.3 Regge trajectories}

 As an effective phenomenological approach, the Regge trajectory~\cite{a503,a504,a501,a502} can help ones to predict the evolution trend of hadron mass spectra. In turn, it could deepen our understanding of the hadron structure by testing the universality of the Regge theory.

 In this paper, the following definition for the $(J, M^{2})$ Regge trajectories is used.
 \begin{eqnarray}
M^{2}=\alpha J+ \beta,
\end{eqnarray}
where $\alpha$ and $\beta$ are the slope and intercept. In Fig.7, the Regge trajectories are plotted in the $(J, M^{2})$
plane based on the calculated mass spectra.
 The fitted slopes and intercepts of the Regge trajectories are given in Table V.

  As shown in Fig.7, the group of $\Xi_{bc}^{'}(NP)$ with the natural parity of $(-1)^{J-1/2}$  is composed of $S(\frac{1}{2}^{+})_{j=1}$, $P(\frac{3}{2}^{-})_{j=1}$, $D(\frac{5}{2}^{+})_{j=2}$, $F(\frac{7}{2}^{-})_{j=3}$ and $G(\frac{9}{2}^{+})_{j=4}$ states. The group of $\Xi_{bc}^{'}(UP)$ with the unnatural parity of $(-1)^{J+1/2}$ is composed of $P(\frac{1}{2}^{-})_{j=1}$, $D(\frac{3}{2}^{+})_{j=2}$, $F(\frac{5}{2}^{-})_{j=3}$ and $G(\frac{7}{2}^{+})_{j=4}$ states.
  For $\Xi_{bc}^{'}$ family, the remaining states in Table I can also be put into these lines, because their mass values are very near those states with the same $L(J^{P})$. The situation is similar for $\Xi_{bc}$, $\Omega_{bc}^{'}$ and $\Omega_{bc}$ families.
From Fig.7, one can see that most of the data points fall on the trajectory lines.

By comparing the slope values in Table V with those of the double-charm and -bottom baryons~\cite{a1055,a1056}, we find a phenomenon: The slopes of the two lines with $n=1$ and $n=3$ are roughly the same. However, the slope of the line with $n=2$ is different from those of the other two lines, and this difference becomes bigger with increasing the total mass of the heavy quark pair.
\begin{figure}[htbp]
\begin{center}
\includegraphics[width=0.9\textwidth]{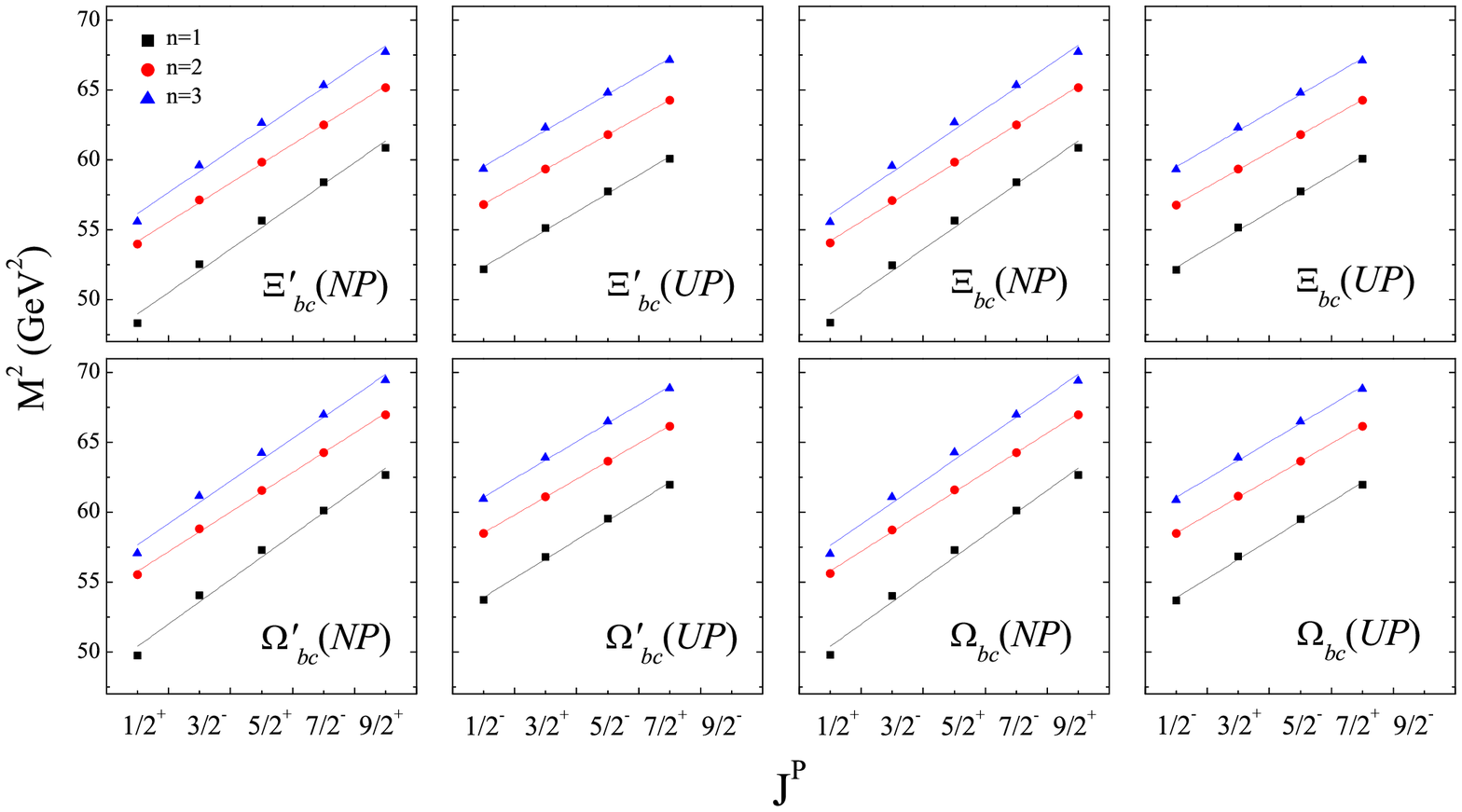}
\end{center}
\caption{(Color online)$(J, M^{2})$ Regge trajectories for the $\Xi_{bc}^{'}$ ($\Xi_{bc}$) and $\Omega_{bc}^{'}$ ($\Omega_{bc}$) families and $M^{2}$ is in GeV$^{2}$. $NP$ denotes the natural parity, and $UP$ the unnatural parity.}
\end{figure}

\begin{table*}[htbp]
\begin{ruledtabular}\caption{Fitted values of the slope ($\alpha$) and intercept ($\beta$) of the Regge trajectories for the $\Xi_{bc}^{'}$ ($\Xi_{bc}$) and $\Omega_{bc}^{'}$ ($\Omega_{bc}$) families.}
\begin{tabular}{c |c c| c c }
Trajectory & \ $\alpha$(GeV$^{2}$) & \ $\beta$(GeV$^{2}$) & \ $\alpha$(GeV$^{2}$) & \ $\beta$(GeV$^{2}$)   \\ \hline
& \multicolumn{2}{c|}{$\Xi_{bc}^{'}$}   &\multicolumn{2}{c}{$\Omega_{bc}^{'}$}   \\ \hline
$n=1(NP)$ &\ $3.099\pm0.191$ &\ $47.409\pm0.548$      &\ $ 3.193\pm0.198 $ &\ $48.800\pm0.568$ \\
$n=2(NP)$  & \  $2.780\pm0.059$ & \ $52.778\pm0.170$  & \ $2.832\pm0.065$  & \ $54.354\pm0.187$  \\
$n=3(NP)$ &\ $3.005\pm0.179$ & \ $54.660\pm0.515$       &\ $3.054\pm0.181$ &\ $56.129\pm0.520$   \\\hline
$n=1(UP) $  & \  $2.632\pm0.097$ & \ $51.014\pm0.222$  & \ $2.739\pm0.102$  & \ $52.540\pm0.234$   \\
$n=2(UP) $  & \  $2.483\pm0.009$ & \ $55.595\pm0.020$  & \ $2.556\pm0.016$  & \ $57.256\pm0.037$  \\
$n=3(UP)$  & \  $2.581\pm0.097$ & \ $58.237\pm0.222$  & \ $2.637\pm0.098$  & \ $59.776\pm0.225$   \\\hline
& \multicolumn{2}{c|}{$\Xi_{bc}$}   &\multicolumn{2}{c}{$\Omega_{bc}$}   \\ \hline
$n=1(NP)$ &\ $3.097\pm0.183$ &\ $47.415\pm0.526$      &\ $ 3.193\pm0.191 $ &\ $48.796\pm0.550$ \\
$n=2(NP)$  & \  $2.770\pm0.048$ & \ $52.812\pm0.137$  & \ $2.823\pm0.055$  & \ $54.379\pm0.159$  \\
$n=3(NP)$ &\ $3.021\pm0.184$ & \ $54.600\pm0.527$       &\ $3.068\pm0.187$ &\ $56.074\pm0.537$   \\\hline
$n=1(UP) $  & \  $2.643\pm0.107$ & \ $50.984\pm0.244$  & \ $2.754\pm0.113$  & \ $52.496\pm0.258$   \\
$n=2(UP) $  & \  $2.495\pm0.020$ & \ $55.563\pm0.046$  & \ $2.562\pm0.024$  & \ $57.237\pm0.054$  \\
$n=3(UP)$  & \  $2.588\pm0.110$ & \ $58.210\pm0.252$  & \ $2.644\pm0.111$  & \ $59.749\pm0.255$   \\
\end{tabular}
\end{ruledtabular}
\end{table*}

\subsection*{3.4 Shell structure of the mass spectra }

 The spectral structures of the $\Xi_{bc}^{'}$ ($\Xi_{bc}$) and $\Omega_{bc}^{'}$ ($\Omega_{bc}$) baryons with $L\leq2$ are presented in Figs.8 and 9, respectively.
  As shown in Fig.8, there are 3 members of the $1S$-wave states for the $\Xi_{bc}^{'}$ and $\Xi_{bc}$ families. The calculated masses are 6952 MeV for $1S(\frac{1}{2}^{+})_{j=1}$ state, 6955 MeV for $1S(\frac{1}{2}^{+})_{j=0}$ state, and 6980 MeV for $1S(\frac{3}{2}^{+})_{j=1}$ state, respectively. One can see that the masses of these three states are relatively close, and the $1S(\frac{1}{2}^{+})_{j=0}$ state of $\Xi_{bc}$ lies between the $1S(\frac{1}{2},\frac{3}{2})_{j=1}$ doublet states of $\Xi_{bc}^{'}$. This suggests that there is a certain difficulty in the identification of the three $1S$ states in experiment. On the other hand, good news is that there lies a big gap (about 240 MeV) between the $1S$ and $1P$ sub-shells as shown in Fig.8. This implies the experimental measurement of the $1S$ states for the $\Xi_{bc}^{'}$ and $\Xi_{bc}$ baryons could be done cleanly. The same is true for $\Omega_{bc}^{'}$ and $\Omega_{bc}$ baryons as shown in Fig.9, where the calculated masses are 7053 MeV for $1S(\frac{1}{2}^{+})_{j=1}$ state, 7055 MeV for $1S(\frac{1}{2}^{+})_{j=0}$ state and 7079 MeV for $1S(\frac{3}{2}^{+})_{j=1}$ state, respectively.

At last, the calculated masses of the $1S$ states in this work are compared with those given by some other theoretical methods as shown in Table VI.
From Table VI, one can see that the masses by the other methods are mainly distributed in the range of 6800$\sim$7100 MeV for $\Xi_{bc}^{'}$ and $\Xi_{bc}$ families, and 6900$\sim$7200 MeV for $\Omega_{bc}^{'}$ and $\Omega_{bc}$ families.
With the above masses, the average value for each $1S$ state is calculated. Table VI shows that our calculated masses are close to the average values in general.

\begin{figure}[htbp]
\begin{center}
\includegraphics[width=0.8\textwidth]{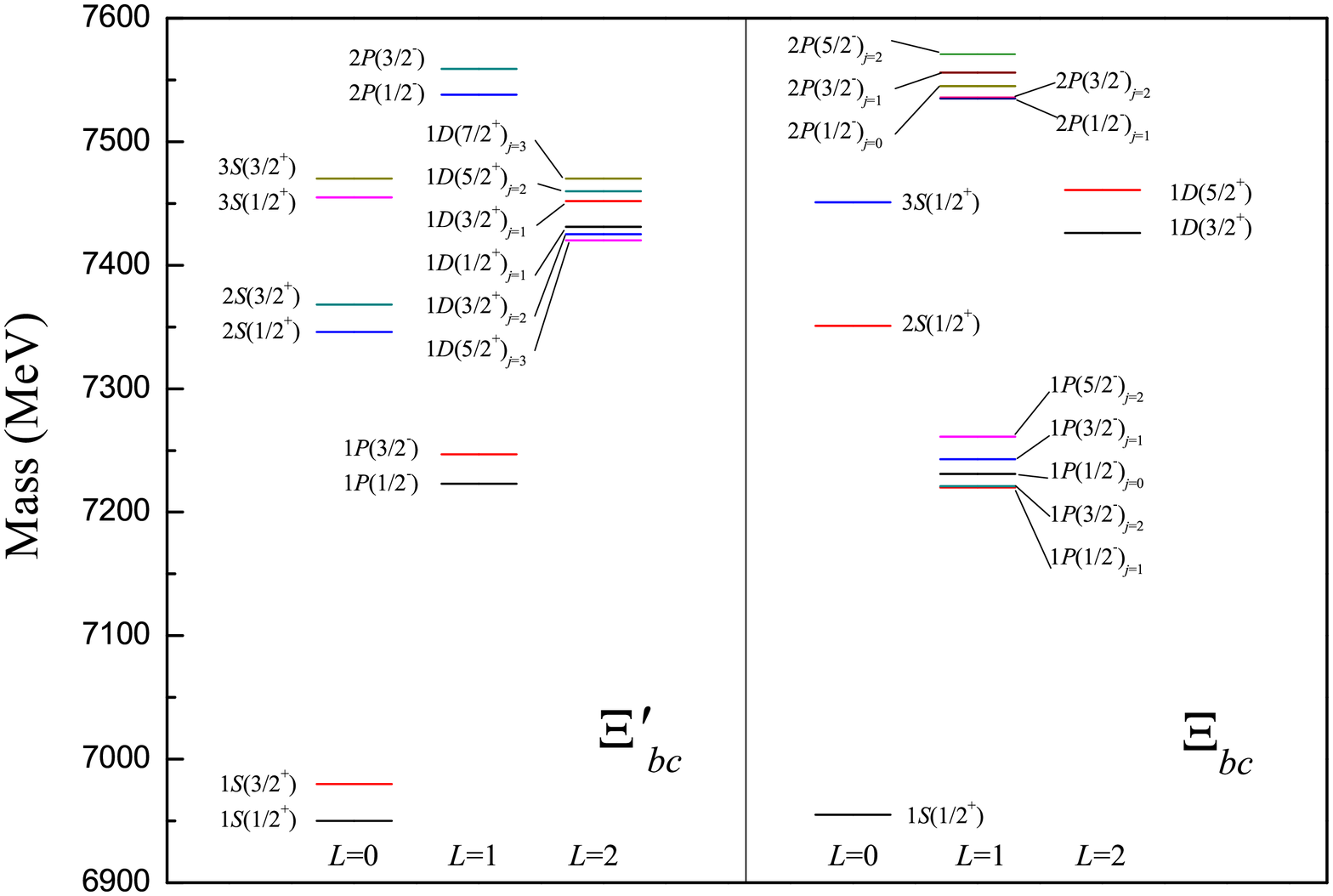}
\end{center}
\caption{(Color online)Shell structure of the $\Xi_{bc}^{'}$ ($\Xi_{bc}$) family.}
\end{figure}

\begin{figure}[htbp]
\begin{center}
\includegraphics[width=0.8\textwidth]{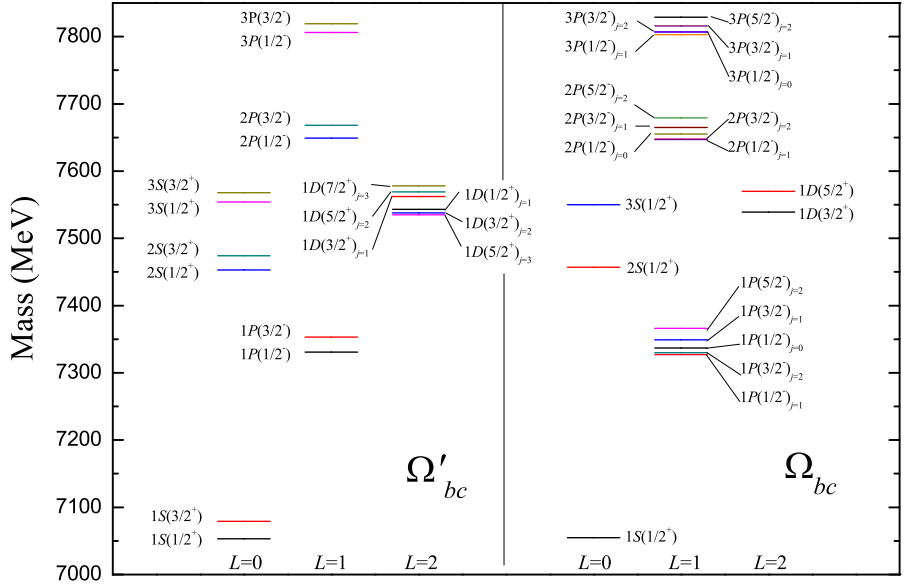}
\end{center}
\caption{(Color online)Same as Fig.8, but for the $\Omega_{bc}^{'}$ ($\Omega_{bc}$) family.}
\end{figure}

\begin{table*}[htbp]
\begin{ruledtabular}\caption{The predicted masses (in MeV) of $1S$ states in this work and some other theoretical methods. The average values ($\bar{m}$) are calculated based on the mass values referenced here. The difference values in the brackets are calculated from $\bar{m}$. }
\begin{tabular}{c c c c c  c c c c c}
 State & $1S(\frac{1}{2}^{+})$ ($\Xi_{bc}^{'}$)&   $1S(\frac{3}{2}^{+})$ ($\Xi_{bc}^{'*}$) &   $1S(\frac{1}{2}^{+})$ ($\Xi_{bc}$) & $1S(\frac{1}{2}^{+})$ ($\Omega_{bc}^{'}$)&   $1S(\frac{3}{2}^{+})$ ($\Omega_{bc}^{'*}$) &   $1S(\frac{1}{2}^{+})$ ($\Omega_{bc}$) \\ \hline
    ~\cite{a1033}&  6948(35) &  6973(-1) & 6922(-9)  &  7047(27) & 7066(-8)  &  7011(-15) \\
  ~\cite{a1036}&  6880(-33) &  6980(6) & 6970(39)  &  6960(-60) & 7060(-14)  &  7050(24) \\
  ~\cite{a1038}&  6934(21) &     - & -     &  7033(13)  & -     &  -   \\
  ~\cite{a1040}&  6805(-108) &  6835(-139) & 6787(-144)  &  6906(-114) & 6930(-144)  &  6893(-133) \\
  ~\cite{a1041}&  6953(40) &  7044(70) & 7015(84)  &  7064(44) & 7142(68)  &  7116(90) \\
  ~\cite{a1042}&  -    &  -    & 6930(-1)  &  -    & -     &  7017(-9) \\
  ~\cite{a1044}&  6890(-23) &  6930(-44) & 6930(-1)  &  7010(-10) & 7040(-34)  &  7040(14) \\
  ~\cite{a1046}&  6958(45) &  6991(17) & -     &  7137(117) & 7170(96)  &  -    \\
  ~\cite{a1049}&  6915(2) &  7003(29) & -     &  -    & -     &  -    \\
  ~\cite{a1050}&  6820(-93) &  6900(-74) & 6850(-81)  &  6930(-90) & 7000(-74)  &  6970(-56) \\
  ~\cite{a1051}&  6800(-113) &  6850(-124) & 6870(-61)  &  6980(-40) & 7020(-54)  &  7050(24) \\
  ~\cite{a1053}&  6950(37) &  7020(46) & 7000(69)  &  7050(30) & 7110(36)  &  7090(64) \\
  ~\cite{a1026}&  7014(101) &  7064(90) & 7037(106)  &  -    & -     &  -    \\
  ~\cite{a1028}&  6988(75) &  7083(109) & -     &  7103(83) & 7200(126)  &  -    \\
  ~\cite{a1029}&  6914(1) &  -    & 6933(2)  &  -    & -     &  -    \\
  ~\cite{a1032}&  6920(7) &  6986(12) & -     &  -    & -     &  -    \\\hline
  $\bar{m}$            &  6913 &  6974 & 6931  &  7020 & 7074  &  7026    \\ \hline
  Our          &  6952(39) &  6980(6) & 6955(24)  &  7053(33) & 7079(5)  &  7055(29) \\
\end{tabular}
\end{ruledtabular}
\end{table*}

\section*{IV. Conclusions}

In this work, by using the relativistic quark model and the ISG method, we investigate the bottom-charm baryon spectra systematically. In the $\rho$-mode, we obtain the mass spectra of the $\Xi_{bc}^{'}$ ($\Xi_{bc}$) and $\Omega_{bc}^{'}$ ($\Omega_{bc}$) families.
The related r.m.s. radii and quark radial probability density distributions are investigated as well, from which we learn more about the structure of bottom-charm baryons.

Based on the obtained mass spectra, we construct successfully the Regge trajectories in the $(J,M^{2})$ plane. We find the slopes of the lines with $n=2$ differ from those of the other lines with $n=1$ and $n=3$, and the difference changes regularly with increasing total mass of the two heavy quarks.

At last, the mass spectral structures of the $\Xi_{bc}^{'}$ ($\Xi_{bc}$) and $\Omega_{bc}^{'}$ ($\Omega_{bc}$) families are presented. We analyze the features of the mass spectral structures, and discuss the difficulty and opportunity of the experimental measurements for the $1S$ states in $\Xi_{bc}^{'}$ ($\Xi_{bc}$) and $\Omega_{bc}^{'}$ ($\Omega_{bc}$) families. We also compare our calculated masses of the $1S$ states with those given by some other theoretical methods. It turns out that our results are close to their average values in general.

\section*{Acknowledgements}

 This research is supported by the Central Government Guidance Funds for Local Scientific and Technological Development of China (No. Guike ZY22096024), the National Natural Science Foundation of China (Grant No. 11675265), the Continuous Basic Scientific Research Project (Grant No. WDJC-2019-13), and the Leading Innovation Project (Grant No. LC 192209000701).

\end{document}